\documentclass[aps,prb,twocolumn,showpacs,amsmath,amssymb]{revtex4}

\usepackage{graphicx}
\usepackage{bm}

\usepackage{color}

\begin{document}

\title{Magnetic phase diagram of the coupled triangular spin tubes for CsCrF$_4$}

\author{Kouichi Seki}
\affiliation{Graduate School of Science and Technology, Niigata University, Niigata 950-2181, Japan}
\author{Kouichi Okunishi}
\affiliation{Department of Physics, Niigata University, Niigata 950-2181, Japan}

\date{\today}

\begin{abstract}
Using Monte Carlo simulations, we explore the magnetic phase diagram of the triangular spin tubes coupled with a ferromagnetic inter-tube interaction for CsCrF$_4$.
A planar structure of the coupled tubes is topologically equivalent to the Kagom\'e-triangular lattice, which induces nontrivial frustration effects in the system.
We particularly find that,  depending on the inter-tube coupling,  various ordered phases are actually realized, such as incommensurate order, ferromagnetic order, and Cuboc order, which is characterized by the non-coplanar spin structure of the twelve sublattice accompanying the spin chirality breaking.
We also discuss a relevance of the results to recent experiments of CsCrF$_4$.
\end{abstract}


\maketitle

\section{Introduction}

Recently, a triangular spin tube has attracted much interest, where its geometrical frustration and quasi-one-dimensionality cooperatively induce exotic magnetic behaviors.
Indeed, theoretical investigations of the $S=1/2$ quantum spin tube\cite{sakai2010quantum} have revealed various interesting properties such as gapful ground state\cite{Kawano,Cabra,Okugap,Fouet,Arikawa,Sakai,Lajko}, field induced chirality order\cite{Sato, Oku}, etc.
Moreover, extensive theoretical researches have been performed for various quantum spin tubes, such as integer-spin tubes\cite{Charrier},  $S=3/2$ triangular tubes\cite{Nishimoto, Fuji}, and four-leg tube\cite{Plat}.
Also, the triangular spin tube has been a target of intensive experimental studies.
For example,  several experiments on [(CuCl$_2$tachH)$_3$Cl]Cl$_2$, which is a $S=1/2$ spin tube consisting of alternating triangles along the tube direction, clarified various characteristic behaviors originating from the tube structure\cite{Schnack, Ivanov}. 
Moreover, straight-type spin tubes CsCrF$_4$ and $\alpha$-KCrF$_4$ have been recently synthesized, which are respectively based on equilateral and non-equilateral triangles\cite{CsCrF4_Babel_1978, CsCrF4_Manaka_2009,CsCrF4_Manaka_2011, manaka2012electron}.
These compounds interestingly  provide essential information about the shape dependence of the unit triangle in the spin tubes.

In CsCrF$_4$,  Cr$^{3+}$ ions having $S=3/2$ spin form a rigid equilateral triangular tube (Fig. 1), where dominant exchange couplings are antiferromagnetic and an inter-tube coupling is estimated to be basically very small.
Accordingly, no anomaly associated with a phase transition was observed by bulk measurements and  ESR experiments down to $T=1.5$K.\cite{CsCrF4_Manaka_2009,CsCrF4_Manaka_2011, manaka2012electron}
On the other hand, a recent experiment of  AC susceptibility observed anomalous slow dynamics suggesting a magnetic long-range order below 4K\cite{CsCrF4_Manaka_2013}.
In addition, a very recent neutron diffraction experiment suggests that this magnetic order is inconsistent with a naive $120^\circ$ structure due to the conventional triangle lattice.\cite{masuda}
Thus, it is expected that the equilateral-triangle structure and a small but finite inter-tube coupling cooperatively induce a non-trivial magnetic structure  in CsCrF$_4$, which could be indeterminate in the bulk quantities.

In order to analyze the magnetic structure of  CsCrF$_4$, a key observation is that Cr$^{3+}$ has a relatively large spin $S=3/2$, and  a certain spin order is suggested by magnetic diffraction peaks in the neutron experiment\cite{masuda}.
Thus, we can expect that the magnetic order of CsCrF$_4$ is basically described by the classical Heisenberg model defined on the triangular tube lattice.
As will be depicted in Figs. 1 and 2, moreover, the inter-tube coupling in the $ab$-plane has the same lattice topology as the Kagom\'e-triangular lattice,\cite{kagome_trianglar_Ishikawa_2014} although the exchange coupling along the $c$-axis is dominant in the spin tube.
Thus, the lattice structure of  CsCrF$_4$ involves the frustration effect even for the ferromagnetic inter-tube coupling.

Of course,  the  Heisenberg model on the planar lattice has no magnetic long-range order at a finite temperature.
In the present coupled tubes,  however, the three-dimensional(3D) couplings possibly stabilize a peculiar spin fluctuation originating from the Kagom\'e-triangular structure.
Using Monte Carlo (MC) simulations, in this paper,  we investigate finite-temperature phase transitions of the classical Heisenberg model on the triangular spin tubes with the inter-tube interaction.
In particular,  we find that the twelve-sublattice spin structure ---Cuboc order--- in the $ab$-plane emerges in the small ferromagnetic inter-tube coupling regime.
The Cuboc order, which  was originally proposed for the ground state of the planar Kagom\'e lattice model with the next-nearest-neighbor interaction, is characterized by a non-coplanar spin structure with the triple-$\bm q$ wave vectors\cite{cuboc_Domenge, domenge2008chirality,Messio}.
In the spin tubes, this Cuboc order can be stabilized to be the 3D long-range order by the strong leg coupling of the tube at a finite temperature.
We also discuss nature of the transitions for the Cuboc phase, as well as incommensurate and ferromagnetic phases, depending on the inter-tube coupling.
Finally  we discuss the relevance to the CsCrF$_4$ experiments.

This paper is organized as follows.
In Sec. II, we explain details of the model and the possible orders.
In Sec. III,  we describe details of MC simulations and definitions of order parameters.
In Sec. IV, we present results of  MC simulations and summarize the phase diagram with respect to the inter-tube coupling.
We also mention universalities of the phase transitions.
In Sec. V, we summarize the conclusion and discuss the relevance to the CsCrF$_4$ experiments.

\section{Model and orders}

\begin{figure}
\centering
\includegraphics[width=1\linewidth]{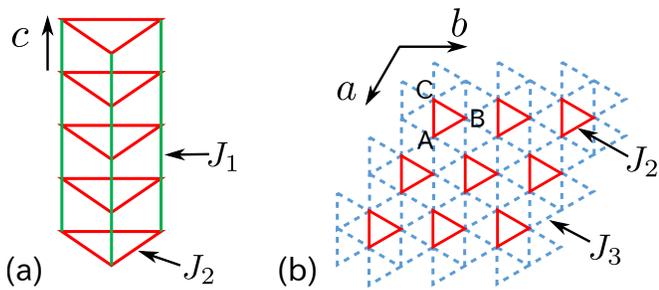}
\caption{
(Color online) Lattice structure of triangular tubes with an inter-tube interaction.
(a) A triangular spin tube, where $J_1$ is the dominant antiferromagnetic coupling in the tube direction ($c$-axis) and $J_2$ denotes the antiferromagnetic interaction in the unit triangle.
(b) The lattice structure in the $ab$-plane. 
The triangles of the solid lines correspond to the $J_2$ coupling in the spin tubes and  the triangles of the broken lines denote the ferromagnetic inter-tube interaction  $J_3$.
\label{fig1}}
\end{figure}

As in depicted in Fig. \ref{fig1}, a bundle of triangular spin tubes in CsCrF$_4$ runs in the $c$-axis direction and these tubes with the inter-tube coupling cover the triangular lattice in the $ab$-plane.
We thus consider the classical Heisenberg model on the stacked triangular lattice, which reads
\begin{eqnarray}
\mathcal H&=&J_1\sum_{\langle i,k\rangle_{c}}\bm S(\bm r_i)\cdot \bm S(\bm r_k)+J_2\sum_{\langle i,j\rangle_\triangle}\bm S(\bm r_i)\cdot \bm S(\bm r_j)\nonumber \\
   && +J_3\sum_{\langle\langle i,j\rangle\rangle}\bm S(\bm r_i)\cdot \bm S(\bm r_j)
\label{Hamiltonian}
\end{eqnarray}
where  $\bm r_i$ represents the position vector of site $i$, and $\bm S(\bm r_i)=S_i^x\bm e_x+S_i^y\bm e_y+ S_i^z \bm e_z $ with $|\bm S_i|=1$ denotes the vector spin at the $i$th site.
Note that  $\bm e_{\alpha}$ ($\alpha\in x,y,z$)  indicates the unit vector in the spin space, while the primitive lattice translation vectors are represented as $\bm a, \bm b$ and $\bm c$ with $|\bm a|=|\bm b|=|\bm c|=1$ (Fig. \ref{fig1}).
Moreover,  $\langle i,k\rangle_{c}$ denotes sum for the nearest-neighbor spins along the $c$-axis, $\langle i,j\rangle_{\triangle}$ indicates sum of spin pairs in the unit triangle, and $\langle\langle i,j\rangle\rangle$ runs over spin pairs of the inter-tube couplings in the $ab$-plane. 
In this paper, we basically assume the antiferromagnetic  interactions $J_1 \ge J_2 > 0$ in the spin tube and the ferromagnetic inter-tube interaction $J_3<0$.
Note that the LDA+U calculation gives $J_1/J_2\simeq 2.0 $ and  $J_3/J_2\simeq 0$ with $J_2\simeq 20$K for CsCrF$_4$\cite{Koo}.

In analyzing possible ordering of the coupled-tube model, an important point is that the dominant coupling $J_1$ along the tube direction does not cause any frustration.
Thus, the staggered pattern of the spin order formed in the $ab$-plane is realized in the $c$-axis direction.
This implies that the low-temperature spin structure is essentially attributed to the frustrating interactions in the $ab$-plane, which we will actually justify with  MC simulations in the next section. 
In the following, we therefore  assume the staggered order in the $c$-axis direction, and concentrate on the spin structure in the $ab$-plane. 

\begin{figure}[tb]
\includegraphics[width=1\linewidth]{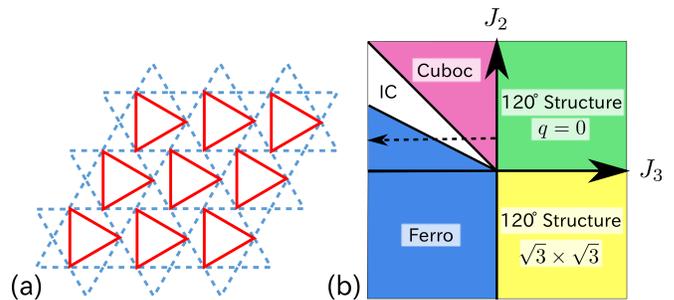}
\caption{
(Color online) (a) Topology of the coupled spin tube lattice in the $ab$-plane  is equivalent to the Kagom\'e-triangular lattice.
(b) The ground state phase diagram of the classical Heisenberg model on the coupled triangular spin tubes.
The left-going arrow with a broken line  shows the direction of $J_3$ parameter corresponding to the finite-temperature phase diagram in Sec. IV.
\label{fig2}}
\end{figure}

As in Fig. \ref{fig2}(a), the planar structure of the model is topologically  equivalent to the  Kagom\'e-triangular lattice, where $J_3$ corresponds to the nearest-neighbor coupling competing with the next-nearest-neighbor interaction $J_2$\cite{kagome_trianglar_Ishikawa_2014}.
Then, a candidate of the ground-state order is classified by Fourier transformation of the Hamiltonian.
Defining the  unit cell as a  triangle of the $J_2$ coupling, we have
\begin{equation}
{\cal H} = \frac{1}{2}\sum J_{\alpha\beta}(\bm q) \bm S_{-\bm q,\alpha}\cdot \bm S_{\bm q, \beta} 
\end{equation}
where ${\bm S}_{\bm q, \alpha}\equiv\frac{1}{\sqrt{N}}\sum_{\bm r} e^{-i\bm q \cdot \bm r}\bm S_{\alpha}(\bm r)$.
Here,  $\bm r$ represents the position of a unit triangle, $\alpha\in \{A,B,C\}$ indicates the sublattice index in the unit triangle, and $N$ is the total number of the unit triangles in the system. 
In addition,  $J_{\alpha\beta}(\bm q)\equiv\sum_{\alpha,\beta} e^{-i\bm q \cdot \bm r_{\alpha\beta}}J_{\alpha\beta}$, where $\bm r_{\alpha\beta}$ and $J_{\alpha\beta}$ respectively denote the relative vector of a spin pair and the corresponding coupling associated with spins in the unit cell.
Note that  the wavevector  $\bm q$ runs over the 1st Brillouin zone in the $ab$ plane.
By determining the lowest energy state of $J_{\alpha\beta}(\bm q)$, we have the ground-state phase diagram of Eq. (\ref{Hamiltonian}) in the $J_2$-$J_3$ plane (Fig.\ref{fig2}(b)), which is equivalent to that obtained in Ref. [\onlinecite{kagome_trianglar_Ishikawa_2014}].

In Fig. \ref{fig2}(b), CsCrF$_4$ is located nearby $J_3\simeq 0$ and  $J_2>0$.
If $J_3$ is antiferromagnetic, Eq. (\ref{Hamiltonian}) is a triangular lattice antiferromagnet for which the ground state is the $120^\circ$ structure.
However, the neutron diffraction experiment suggested that the order of  CsCrF$_4$ is not explained by a naive $120^\circ$ structure\cite{masuda}.
We thus discuss the negative  $J_3( < 0)$ region, where the nontrivial exotic order phase actually emerges;
In $-J_3 \le J_2$ region, particularly, the minimum of $J_{\alpha\beta}(\bm q)$ located at  M point in the Brillouin zone, where the non-coplanar order with the twelve sublattice that is called "Cuboc" order can be stabilized.
As $-J_3$ increases, the incommensurate order appears in the $J_2<-J_3<2J_2$ region, and finally the ferromagnetic order becomes stable for $-J_3 \ge 2J_2$.
This ground-state phase diagram suggests that,  in CsCrF$_4$,  the Cuboc state can be stabilized  to be the 3D long-range order at a finite temperature by the tube-leg coupling $J_1$, even if the amplitude of the inter-tube interaction $J_3$ is very small.

\begin{figure}
\centering
\includegraphics[width=1\linewidth]{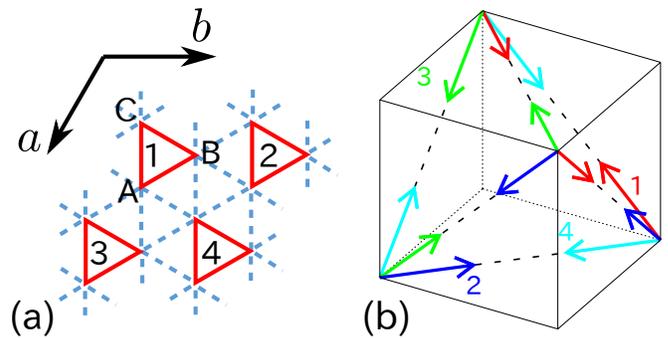}
\caption{
(Color online) (a) The extended unit cell for the Cuboc order with the twelve sublattice structure.
The label \{A, B, C\} indicates each vertex in the unit triangle and the number \{1,2,3,4\} represents the label of a triangle in the extended unit cell.
(b) The arrow with the numbers shows the spin directions forming  $120^\circ$ structure in the corresponding unit triangle.
The four tilting  $120^\circ$-structures in the extended unit cell form the tetrahedron represented by the dashed lines.
\label{fig3}}
\end{figure}

Here, we briefly summarize the essential property of the Cuboc order in $0<-J_3< J_2$, which was originally found in the  $J_1-J_2$ Heisenberg model on the Kagom\'e lattice\cite{cuboc_Domenge,domenge2008chirality,Messio}.
The Cuboc order has a non-coplanar spin structure accompanying the spontaneous symmetry breaking of the lattice translation.
Fig. \ref{fig3}(a) shows the extended unit cell for the Cuboc order in the $ab$-plane, where we assign a number to each of four triangles, for later convenience.
Then, the strong antiferromagnetic coupling $J_2$ basically imposes the 120$^\circ$ structure in each triangle.
Then, a significant point is that the 120$^\circ$-structure planes can relatively tilt among the four triangles, so as to reduce the energy due to the frustrating $J_3$ interaction.
Gluing the four tilting triangles of the 120$^\circ$ structure, we obtain the tetrahedron where the twelve spins in the extended unit cell are attached.
As shown in Fig. \ref{fig3}(b), the three spins on the unit triangle of the original lattice  are mapped into the vertices of the corresponding triangle on the tetrahedron, where the vector-spin chiralities sitting on the four 120$^\circ$ planes point the radial direction from center of the tetrahedron.
In this sense, the vector-spin chirality associated with the tetrahedron can be a good order parameter of the non-coplanar Cuboc spin structure.
Note that, if the spin vectors are arranged at the origin of the spin space, we have the same schematic diagram as in Refs.[\onlinecite{cuboc_Domenge,domenge2008chirality,Messio}].

For the Cuboc order, the magnetic propagation vectors interestingly have a triple $\bm q$ structure in the $ab$-plane, reflecting the above characteristic spin configuration.
Let us write the sublattice spin in the unit cell at position $\bm r$ as $\bm S_\alpha(\bm r)$, where $\alpha\in \{A, B, C\}$ is the sublattice index in the unit triangle.
Then, the Cuboc order can be explicitly written as  
\begin{align}
\bm S_A(\bm r)&=\cos\left(\frac{1}{2}\bm q_a\cdot \bm r\right)\frac{\bm e_x}{\sqrt 2}-\cos\left(\frac{1}{2}\bm q_\gamma\cdot \bm r\right)\frac{\bm e_y}{\sqrt 2}\nonumber\\
\bm S_B(\bm r)&=\cos\left(\frac{1}{2}\bm q_\gamma\cdot \bm r\right)\frac{\bm e_y}{\sqrt 2}-\cos\left(\frac{1}{2}\bm q_b\cdot \bm r\right)\frac{\bm e_z}{\sqrt 2}\nonumber\\
\bm S_C(\bm r)&=\cos\left(\frac{1}{2}\bm q_b\cdot \bm r\right)\frac{\bm e_z}{\sqrt 2}-\cos\left(\frac{1}{2}\bm q_a\cdot \bm r\right)\frac{\bm e_x}{\sqrt 2}\label{equ_cuboc}
\end{align}
where $\bm q_a, \bm q_b$ are the reciprocal lattice vectors of the primitive translation vectors $\bm a, \bm b$, and $\bm q_\gamma \equiv \bm q_a-\bm q_b$.
This triple $\bm q$ is an important feature of the Cuboc phase, and plays an essential role in the analysis of the neutron diffraction experiment.
Finally, we note that, in the $c$-axis direction, the magnetic propagation vector is $\bm q_c/2$, where the simple staggered pattern appears.

\section{Monte Carlo simulation}

In the previous section, we have discussed the ground-state orders of the triangular spin tubes in the $ab$-plane for $-J_3\le J_2$.
So far, investigations of the planar Kagom\'e-triangular-lattice model have clarified that, although there exists no true long-range order of the spin,  the chirality degrees of freedom associated with the Cuboc order induces the $Z_2$ symmetry breaking transition even at a finite temperature\cite{cuboc_Domenge,domenge2008chirality}.
For the coupled-tube system, which contains the full 3D couplings, we can expect the finite temperature transitions associated the Cuboc long-range order,  the incommensurate order, as well as the ferromagnetic order.
In order to address  the nature of these finite-temperature transitions, we perform extensive Monte Carlo simulations for the coupled-tube system of Eq. (\ref{Hamiltonian})

Here, we comment on the notation of the system size.
In the following,  we basically represent the linear dimensions of the system by the number of triangles associated with $J_2$ couplings.
Thus, $L_a$($L_b$) means a number of triangles in the $a$($b$)-axis direction, and $L_c$ denotes length of a tube in the $c$-axis direction.
In this paper, we basically deal with the system of $L_a=L_b=L_c\equiv L$, for which  the total number of spins in the system is $N=3L^3$.

\subsection{Details of simulation}

We  employ the  Wolff's cluster algorithm\cite{wolff1989collective} combined with the Metropolis local update.
Usually, the Wolff algorithm is not efficient for frustrated systems, where  a large cluster containing almost all of spins are often generated.  
However, we find that the coupled spin tubes has no frustration in the $c$-axis, which makes possible cluster growing of an efficient size in the tube-direction.
Thus the Wolff algorithm works very well for the coupled spin tubes.
Note that  the parallel tempering method\cite{hukushima1996exchange} is additionally used in practical computations, if the relaxation to the equilibrium is difficult.
On the basis of the above algorithm, we have performed  MC simulations for the system of $L=8, 16, \cdots , 36$  with the periodic boundary condition.
We particularly explore the  $T - J_3$ phase diagram along the broken line in Fig. \ref{fig2}(b) with $J_2=1.0$ fixed. 
Typical numbers of  MC samples are  $2^{19}\sim 2^{23}$.

\subsection{Order parameters}

As pointed out in the ground-state phase diagram of Fig. \ref{fig2}(b), the coupled triangular spin tubes have the various ordered states.
To classify these ordered states in simulations, we need to define appropriate order parameters.
In particular, the Cuboc order has the non-coplanar spin  structure with the twelve sublattice in the $ab$-plane, which is staggeredly stacked in the $c$-axis direction.
Taking account of this structure, we define the Cuboc sublattice magnetization $\bm m_{\alpha,\beta}$ in a certain  $ab$-plane, where $\alpha\in \{A, B,C\}$  indicates a vertex of a triangle, and $\beta \in \{1,2,3,4\}$ specifies a triangle in the extended unit cell (Fig. \ref{fig3}(a)). 
\begin{equation}
\bm m_{\alpha,\beta}\equiv\frac{1}{N_p}  \sum_{\bm r_\beta}{}' \bm S_{\alpha}(\bm r_\beta) 
\end{equation}
where $\bm r_\beta$ denotes the position of the triangle labeled by $\beta$ in the extended unit cell, and $\sum'_{\bm r_\beta}$ represents the sum with respect to triangles having the same $\alpha$ in the plane.
In addition, the normalization $N_p\equiv L^2/4$ is the  number of the extended unit cell in the $ab$ plane.
Then, the sublattice magnetization order parameter in the entire system is defined as
\begin{equation}
\bm M_{\alpha,\beta} \equiv \frac{1}{L} \sum_{i_c} (-1)^{i_c} \bm m_{\alpha,\beta}
\end{equation}
where $i_c$ denotes index of the tube direction (Fig. \ref{fig1}(a)).

In $-J_3 > 2J_2 $ region,  the ground state of the system is the ferromagnetic ordered state in a certain Kagom\'e-triangular plane .
Thus, we also introduce the ferromagnetic order parameter  as
\begin{equation}
\bm m_{\rm F}\equiv\frac{1}{12}\sum_{\alpha,\beta} \bm m_{\alpha,\beta},
\end{equation}
which detects the uniform magnetization in a Kagom\'e-triangular plane.
Then, this in-plane ferromagnetic order is staggeredly stacked in the $c$-axis direction.
Thus we define the total ferromagnetic order parameter $\bm M_{\rm F}$ as
\begin{equation}
\bm M_{\rm F}\equiv \frac{1}{L}\sum_{i_c} (-1)^{i_c} \bm m_{\rm F}.
\end{equation}

The Cuboc order is basically  detectable by the 12-sublattice magnetization $\bm M_{\alpha\beta}$.
As was mentioned in the previous section, moreover, there is another essential order parameter ---chirality degrees of freedom--- originating from the $120^\circ$ structure on the unit triangle.
For the unit triangle at $\bm r$, we define the  vector spin chirality\cite{Miyashita},
\begin{eqnarray}
\bm \kappa(\bm r)\equiv\frac{2}{3\sqrt 3}\left[\bm S_{A}(\bm r)\times \bm S_B(\bm r)+\bm S_{B}(\bm r)\times \bm S_{C}(\bm r) \right. \nonumber\\
\left. +\bm S_{C}(\bm r)\times \bm S_{A}(\bm r) \right] .
\end{eqnarray}
An important point on the Cuboc order is that the vector chiralities associated with the four triangles in the extended unit cell also have non-coplanar configuration.
Thus, we introduce the sublattice vector chirality order parameter for the Cuboc state as
\begin{equation}
\bm K_\beta\equiv\frac{1}{LN_p}\sum_{i_c}\sum_{\bm r_\beta}{}' \bm \kappa(\bm r_\beta),
\end{equation}
where $\sum_{\bm r_\beta}{}'$ sums up $\bm \kappa$ carrying the sublattice index $\beta$ of the extended unit cell.
Here, note that the vector spin chirality takes the same sign in the $c$-axis direction, although the direction of the spins are alternatingly aligned.

In the Cuboc phase, both of $\bm M_{\alpha,\beta}$  and $\bm K_\beta$  have  finite values, but $\bm M_{\rm F} = 0$.
In the ferromagnetic state, $\bm M_{\rm F}$ is finite, while $\bm K_\beta = 0$.
For the incommensurate phase between the Cuboc and ferromagnetic phases, we do not set up a direct order parameter of the incommensurate spin configuration, because it is very difficult to determine the pitch of the incommensurate oscillation within the system size up to $L=36$.
However, it should be noted that  $\bm M_{\alpha,\beta}=0$  and $\bm K_\beta=0$ is basically seen in the incommensurate phase.

\section{Results}

We present results of MC simulations for typical parameters of the coupled spin tubes.
As mentioned in Sec. II, the parameters corresponding to the CsCrF$_4$ is that $J_1/J_2 \simeq 2$ and $ J_3/J_2\simeq 0$.
In this paper, we fix $J_2=1.0$ and investigate the inter-tube-coupling ($J_3$) dependence for $J_1=1.0$ and  $3.0$, along the arrow with the dotted line in the ground-state phase diagram as in Fig. \ref{fig2}(b).
We then find no qualitative difference between $J_1=1.0$ and $3.0$, so that we show the results for  $J_1=1.0$ below.

\subsection{Phase diagram}

\begin{figure}[ht]
\centering
\includegraphics[width=0.7\linewidth]{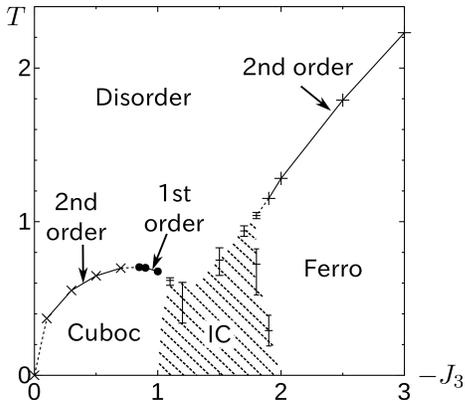}
\caption{
$T$-$J_3$ phase diagram of the coupled triangular spin tubes with $J_1=J_2=1.0$.
The Cuboc order is realized  at finite temperature by the inter-tube interaction.
\label{phase}}
\end{figure}

In Fig. \ref{phase},  we show the final phase diagram in the $T$-$J_3$ plane, before presenting detailed analysis of simulations.
Note that the horizontal axis represents $-J_3$, since we consider the negative $J_3$ region.
In the figure, it is verified that the long-range orders at a finite temperature are consistent with the ground-state phase diagram of the planar Kagom\'e-triangular model in Fig. \ref{fig2}.
For $0<-J_3 \lesssim 1 $, the Cuboc phase is actually realized, and, in $1 \lesssim  -J_3 \lesssim 2$,  the incommensurate spin order appears.
Moreover, the system exhibits the ferromagnetic order in $-J_3 \gtrsim 2$,  where the ferromagnetic coupling is dominant.

The phase boundary of the Cuboc-disorder transition is of second order for $-J_3\lesssim 0.7$, which can be determined with a finite-size-scaling analysis.
The universality of this second-order-transition line possibly belongs to a novel universality class associated with the triple-$\bm q$ structure of the Cuboc order.
A detailed scaling analysis will be presented in the following subsection.
For $0.85\lesssim -J_3 \lesssim 1.1$, however, the Cuboc-disorder transition changes to the first order, where the double peak of the energy histogram is observed.
Note that a precise identification of the transition is difficult around $-J_3\sim 0.8$, where a tricritical point is expected.
In $1.1 \lesssim -J_3 \lesssim 1.8 $,  the incommensurate-disorder transition line is estimated by a peak position of specific heat within finite size systems up to $L=32$,  where the size extrapolation is difficult.
For $-J_3 \gtrsim 1.8 $,  the line of the disorder-ferromagnetic order transition  is of second order, for which the universality class is consistent with the 3D ferromagnetic Heisenberg model.
On the other hand, the transitions between the incommensurate phase and the other ordered phases are expected to be a commensurate-incommensurate type.
These transition lines associated with the incommensurate order are also estimated within finite size results.

\subsection{Cuboc phase}

\begin{figure}[tb]
\centering
\includegraphics[width=1\linewidth]{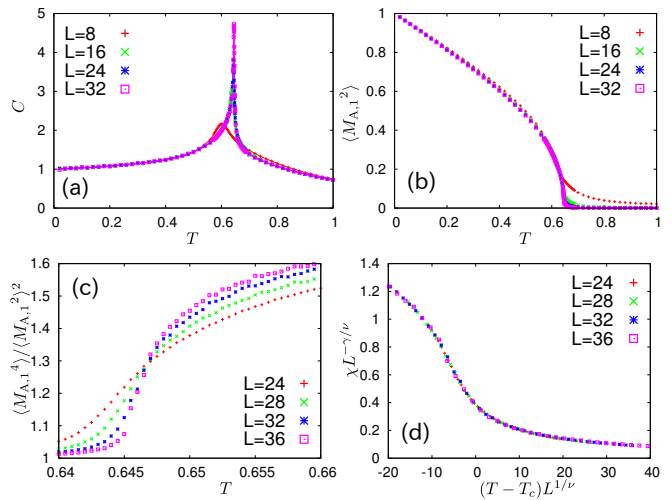}
\caption{
(Color online) Results of MC simulations for $-J_3 = 0.5$.
(a) Specific heat $C$.
(b) Mean-square average of the Cuboc sublattice  magnetization $\langle {\bm M_{\rm A, 1}}^2 \rangle $.
(c) Binder cumulant for $\bm M_{\rm A, 1}$.
(d) Finite-size-scaling plot for the Cuboc order parameter, which yields $T_c=0.64721(4)$, $\nu=0.435(4)$ and $\gamma=0.90(2)$.
\label{fig_cuboc}}
\end{figure}

\begin{figure}[t,b]
\includegraphics[width=1\linewidth]{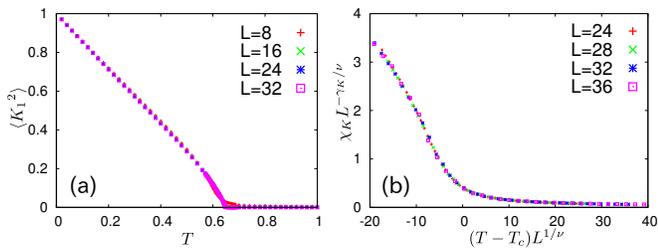}
\caption{
(Color online) Results of the vector spin chirality for $-J_3 = 0.5$.
(a) Mean-square average of the sublattice vector spin chirality $\langle {{\bm  K}_{1}}^2\rangle $.
(b) Finite-size-scaling plot for  $\langle {{\bm  K}_{1}}^2\rangle $, which yields best fit values: $T_c=0.64742(4)$, $\nu=0.433(8)$ and $\gamma_K=0.57(2)$.
\label{fig_chirality_cuboc}}
\end{figure}

Let us begin with detailed analysis of the Cuboc phase.
In Figs. \ref{fig_cuboc} and \ref{fig_chirality_cuboc}, we present MC results for the Cuboc phase.
Specific heat $C$ and a mean-square average of the Cuboc sublattice magnetization $\langle {\bm M_{{\rm A},1}}^2 \rangle $ are shown in Fig. \ref{fig_cuboc}(a) and (b).
First of all,  the specific heat $C$ has a sharp single peak at $T_c\sim 0.65 $, and the sublattice magnetization ${\bm M}_{{\rm A},1}$ also exhibits the phase transition behavior at the same temperature.
In order to precisely determine the transition point, we calculate the Binder cumulant\cite{binder1981critical} of the Cuboc sublattice magnetization $\langle {\bm M_{{\rm A},1}}^4\rangle / \langle {{\bm M}_{{\rm A},1}}^2\rangle^2$.
The result for $-J_3=0.5$ is shown in Fig. \ref{fig_cuboc} (c),  where the curves for various system sizes cross at $T_c=0.6470(5)$.
This implies that the transition is of second order, prompting us to determine the universality class of the Cuboc transition with a finite-size-scaling analysis. 
Assuming the scaling form for the susceptibility of the Cuboc sublattice magnetization,
\begin{equation}
\chi \equiv LN_p\langle {{\bm M}_{\rm A, 1}} ^2 \rangle/T  \propto L^{\gamma/\nu} \Psi(tL^{1/\nu})
\end{equation}
with $t=(T-T_c)/T_c$, we perform the Bayesian estimation for the critical exponents $\nu$, $\gamma$ and $T_c$.\cite{harada2011bayesian}
Fig. \ref{fig_cuboc}(d) shows the resulting finite-size-scaling plot for $\chi$ with the best-fit value $\nu=0.435(4)$,  $\gamma=0.90(2)$ and $T_c=0.64721(4)$.
Here, we note that $T_c$ is consistent with the result of the Binder cumulant, although it is obtained independently of the Binder-cumulant result.
Taking account of the error originating from choice of the data window, we finally adopt $\nu=0.44(2)$ and $\gamma=0.91(3)$ for the critical exponents of the Cuboc sublattice magnetization.

As mentioned before, the sublattice vector spin chirality  is also another essential order parameter of the Cuboc order.
Fig. \ref{fig_chirality_cuboc} (a) shows  $\langle {{\bm K}_1}^2\rangle $ for  $-J_3=0.5$, where the transition occurs at the same $T_c$ as the sublattice magnetization $ {\bm M}_{{\rm A},1}$.
This behavior is consistent with the observation that the specific heat has a single peak at $T_c$.
Thus, we can expect that the Cuboc magnetization and the chirality degrees of freedom exhibit the simultaneous transition.
We then perform the finite-size-scaling analysis for the chirality susceptibility with $\chi_K \equiv LN_p\langle {{\bm K}_1}^2\rangle/ T \propto L^{\gamma_K/\nu} \Psi_K(tL^{1/\nu }) $. 
In Fig. \ref{fig_chirality_cuboc} (b),  we present the finite-size-scaling plot  with  $T_c=0.64742(4)$, $\nu=0.433(8)$ and $\gamma_K=0.57(2)$, which are also obtained with the Bayesian estimation\cite{harada2011bayesian}.
Taking account of the data-window dependence, we finally identify the exponents associated with the vector spin chirality as $\nu=0.43(2)$ and $\gamma_K=0.55(4)$.
Here, it should be noted that, although no apriori assumption of $T_c$ and $\nu$ was set up in this scaling analysis,  the resulting $T_c$ and $\nu$ are consistent with those for $\chi$, while $\gamma_K$ is clearly different from $\gamma$.
These facts suggest that the singular part of the free energy is scaled with
\begin{equation}
f_s \propto L^{-d}f_s(tL^{1/\nu} , hL^{y}, h_K L^{y_K})
\end{equation}
where $y(=(\frac{\gamma}{\nu}+d)/2)$ and $y_K(=(\frac{\gamma_K}{\nu}+d)/2)$ are the eigenvalues of the linearized-renormalization-group transformation corresponding to the fields conjugated to the Cuboc magnetization and the chirality, respectively. 
Thus we have concluded that the transition of the spin and chirality degrees of freedom  is simultaneous.
Here, the exponents obtained for the Cuboc transition are clearly different from those of the layered triangular lattice antiferromagnet, although the simultaneous transition was also observed\cite{Kawamura, mailhot1994finite, Kawamura_1998, Kawamura_2001}. 
The universality of the Cuboc transition provides a novel class, which might be characterized by the effective chiral Ginzburg-Landau-Wilson theory associated with the triple $\bm q$-structure\cite{Kawamura1988,Kawamura_1990,Kawamura_1998}. However, the detailed analysis of the effective model is an interesting future issue.
Here, we remark that  ${\bm M}_{\alpha,\beta}$ and ${\bm K}_\beta$ are confirmed to be consistent among the all combinations of sublattice indices $\alpha$ and $\beta$. 
On the basis of the analysis above, we have finally drawn the second-order-transition line of the Cuboc phase in Fig. \ref{phase}.

\begin{figure}[b,t]
\centering
\includegraphics[width=1\linewidth]{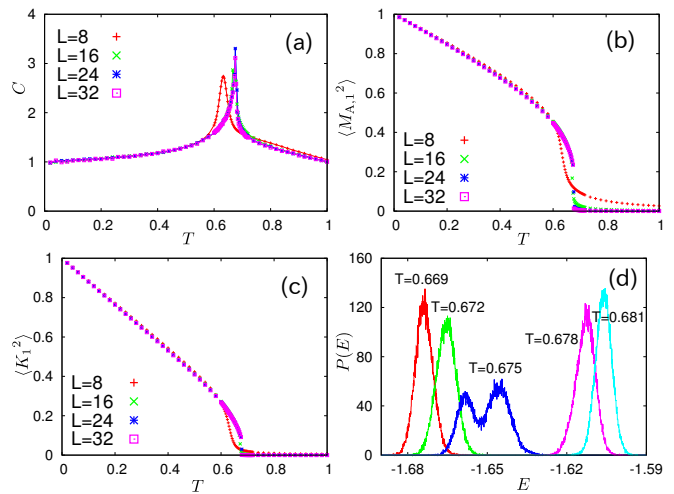}
\caption{
(Color online) Results of MC simulations at $-J_3=1.0$.
(a) Specific heat $C$.
(b) Mean-square average of the Cuboc sublattice  magnetization $\langle {\bm M_{\rm A, 1}}^2 \rangle $.
(c) Mean-square average of the sublattice vector spin chirality $\langle {{\bm  K}_{1}}^2\rangle $.
(d) Energy histogram around the transition point for $L=32$.
The double peak structure emerges at $T_c=0.675$, indicating that the transition is of first-order.
\label{fig_first}}
\end{figure}

We turn to the first-order transition in the region of $0.85  \lesssim  -J_3 \lesssim 1.1 $.
Figure  \ref{fig_first} shows MC results for $-J_3 = 1.0$.
In Fig. \ref{fig_first}(a),  the specific heat $C$ also has a sharp single peak at $T_c\sim 0.675$.
In Fig. \ref{fig_first}(b) and (c), moreover, we can observe that mean-square averages of the Cuboc sublattice magnetization $\langle{{\bm  M}_{{\rm A},1}}^2\rangle $ and of the vector spin chirality $\langle {{\bm K}_1}^2 \rangle$ also  exhibit the phase transition behavior at the same temperature $T_c$, illustrating the simultaneous transition of the spin and chirality degrees of freedom.
In the figures, we can also observe that the both of $\langle {{\bm M}_{{\rm A},1}}^2 \rangle $ and $\langle {{\bm K}_1}^2 \rangle$ for $L=24$ and 32 show small jumps at $T_c$, suggesting that the transition is of first order.
We have computed the energy histogram around $T_c$ to confirm its double peak structure at $T_c=0.675$ (Fig. \ref{fig_first}(d)).
Thus, we conclude that the transition  in $0.85  \lesssim  -J_3 \lesssim 1.1 $ is the first-order.

Finally, we would like to comment on the tricritical point expected around $-J_3 \sim 0.8$.
As varying $J_3$, we have checked that the double peak of the energy histogram appears down to $-J_3=0.85$, while the crossing point of the Binder cumulant emerges up to $-J_3=0.7$.
Thus, the tricritical point is possibly located around $-J_3 \sim 0.8$.
Within the present system size of the MC simulation, however, a precise identification of the tricritical point is difficult. 
The detailed analysis on this respect is a future issue.

\subsection{Ferromagnetic phase}

\begin{figure}[h]
\includegraphics[width=1\linewidth]{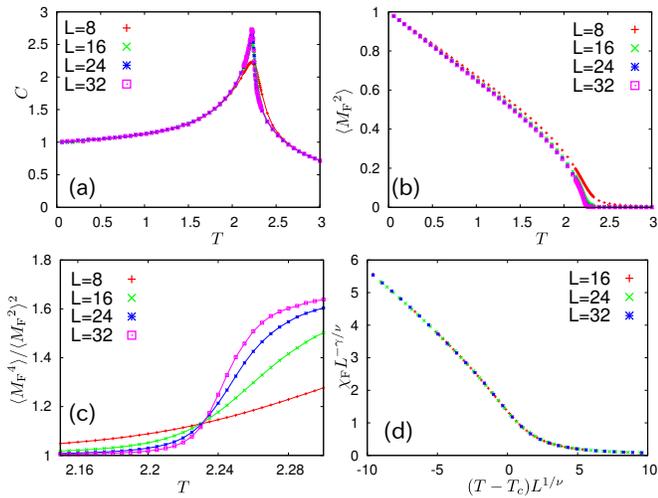}
\caption{
(Color online) Results for $-J_3 = 3.0$.
(a) Specific heat $C$.
(b) Mean-square average of the uniform magnetization $\langle  {{\bm M}_{\rm{F}}}^2\rangle $.
(c) Binder cumulant for ${\bm M}_{\rm{F}}$.
(d) Finite-size-scaling plot for the susceptibility of $\chi_{\rm F}$.
The fitting result is consistent with the 3D Heisenberg  universality class\cite{holm1993critical, nVector_Guillou, 3D_Heisen_Campostrini, 3D_Heisen_Pelissetto}.
\label{fig_ferro}}
\end{figure}

In the $-J_3 \gtrsim 2.0$ region, the ferromagnetic coupling becomes dominant, where the system forms the ferromagnetic order in a Kagom\'e-triangular layer, accompanying  the second-order phase transition.
In Fig. \ref{fig_ferro}, we show results of MC simulations at $-J_3 = 3.0$.
The specific heat $C$ in Fig. \ref{fig_ferro}(a) indicates a peak at $T_c\sim 2.2 $, and  the magnetization $\langle {{\bm M}_{\rm F}}^2 \rangle $ in Fig.  \ref{fig_ferro}(b) also exhibits the phase transition behavior.
Note that the sublattice spin chirality $\langle {{\bm K}_1}^2 \rangle$ is checked to be  always zero in the region of the ferromagnetic phase.

In order to precisely determine $T_c$, we further calculate the Binder cumulant $
\langle {{\bm M}_{\rm{F}}}^4\rangle/\langle {{\bm M}_{\rm{F}}}^2\rangle^2
$.
Its result is shown in Fig. \ref{fig_ferro}(c), where the crossing point appears at $T_c=2.233(3)$.
We also perform the finite-size-scaling  analysis of the susceptibility $\chi_{\rm F}\equiv 12LN_p\langle {{\bm  M}_{\rm{F}}}^2\rangle/T $, using the Bayesian estimation.
The finite-size-scaling plot in Fig. \ref{fig_ferro}(d) well collapses to a scaling function with $T_c=2.235(5)$, $\nu=0.70(2)$ and $\gamma=1.40(5)$.
Note that these exponents are clearly consistent with the 3D ferromagnetic Heisenberg class: $\nu \simeq 0.71$ and $\gamma \simeq1.40$\cite{holm1993critical, nVector_Guillou, 3D_Heisen_Campostrini, 3D_Heisen_Pelissetto}.
Thus, we can conclude that the universality class of the transition for the ferromagnetic order is the conventional 3D ferromagnetic Heisenberg class.

\subsection{Incommensurate phase}

\begin{figure}[t,b]
\centering
\includegraphics[width=1\linewidth]{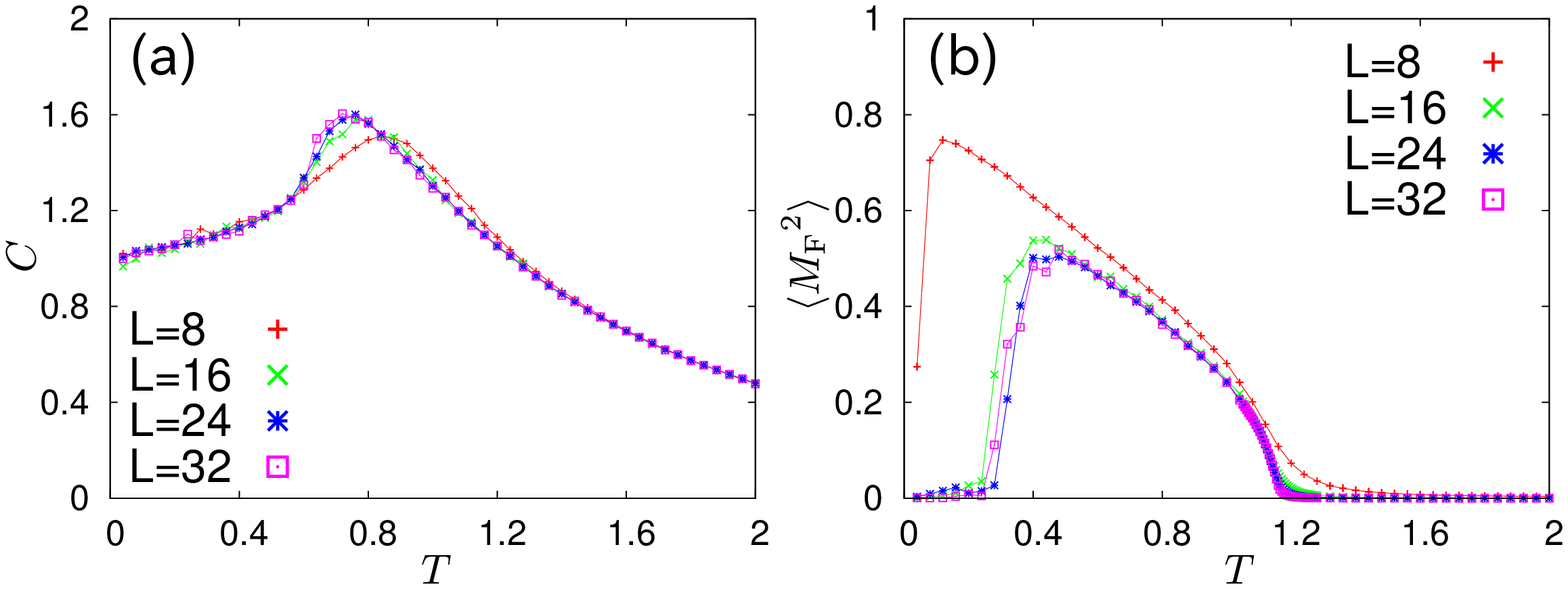}
\caption{
(Color online) Results of MC simulations for the  incommensurate phase.
(a) Specific heat $C$ for  $-J_3 = 1.5$, which shows  a round peak at $T\sim 0.7$.
(b) Mean-square average of the ferromagnetic magnetization $\langle  {{\bm M}_{\rm{F}}}^2\rangle $ at $-J_3 = 1.9$, which abruptly decays in the incommensurate order region.
\label{fig_ic}}
\end{figure}

Let us finally discuss the incommensurate phase in the $ 1.1 \lesssim  - J_3 \lesssim 2 $ region, which is sandwiched between the Cuboc and ferromagnetic phases. 
In Fig. \ref{fig_ic}(a), we show the specific heat $C$ at $-J_3 = 1.5$, where a phase transition is illustrated by a rounded peak of $C$ around $T_c \sim 0.7$.
However, the size extrapolation to extract the bulk behavior is usually difficult for the incommensurate order, where the pitch of the oscillation does not match with the system size.
Thus, we basically estimate the transition temperature by the peak position of the specific heat of $L=32$.

Turning to the transition line between the incommensurate and ferromagnetic phases, we can expect the commensurate-incommensurate type transition, where the wave vector of the order may continuously sift from the $\Gamma$ point toward the M point, following the ground-state phase diagram in Fig. 2(b).
Figure \ref{fig_ic}(b) shows the uniform magnetization $\langle {{\bm M}_{\rm F}}^2\rangle$ for $-J_3 = 1.9$, where we observe that it takes a finite value in $0.35 \lesssim T \lesssim 1.2 $, but rapidly decays below $T\sim 0.35$.
This behavior indicates that the ferromagnetic order appears between $0.35 \lesssim T \lesssim 1.2 $, but it abruptly changes into the incommensurate order in the low-temperature region of  $T\lesssim 0.35$.
We thus define the boundary between ferromagnetic and incommensurate phases as the middle point of the onset and offset of $\langle{{\bm M}_{\rm{F}}}^2\rangle$ for $L=32$.

On the other hand, we  note that the transition line between the incommensurate and Cuboc phases is difficult to estimate by the result within $L=32$.
Thus, the border between the incommensurate and Cuboc phases in Fig. \ref{phase} is just a guide for eyes.

\section{Summary and Discussions}

In this paper, we have investigated phase transitions of the coupled triangular spin tubes associated with CsCrF$_4$.
An essential point is that a two-dimensional section of the coupled tubes forms a Kagom\'e-triangular plane, which drives the system into  exotic orders such as  Cuboc order, incommensurate order and ferromagnetic order.  
In particular, the Cuboc order is  characterized by the twelve sublattice non-coplanar spin structure carrying the triple-$\bm q$ wave vectors, which accompanies the non-coplanar structure of the vector spin chirality as well.
Performing extensive Monte Carlo simulations, we have demonstrated that these phases are actually realized for the coupled tubes with the negative inter-tube coupling $J_3$ at a finite temperature.
The resulting phase diagram was summarized in Fig. \ref{phase}.
Then, a particular finding is that the transition to the Cuboc order in $-J_3 \lesssim 0.7$ is described by the simultaneous second-order transition of the spin and chirality degrees of freedom. 
To our knowledge, the universality of this transition is a novel class characterized by the non-coplanar spin structure accompanying the chirality, which might be described by an effective chilal Ginzburg-Landau-Wilson theory associated with the triple $\bm q$-structure.\cite{Kawamura1988,Kawamura_1990, Kawamura_1998}
While for $0.85 \lesssim-J_3 \lesssim 1.1 $,  the transition is of first order, where the double peak of the energy histogram is confirmed.
However, the analysis of the expected tricritical point is a remaining issue.
On the other hand,  for  $-J_3 \gtrsim 1.9 $, we have confirmed that the ferromagnetic transition belongs to the 3D Heisenberg universality class.

From the experimental view point, CsCrF$_4$ is described by the weak $J_3$ coupling limit of the present model.
The  neutron scattering experiment of CsCrF$_4$ actually suggests that a possible spin order is not a naive 120$^\circ$ structure,\cite{masuda} and thus a finite temperature transition to the Cuboc phase can be expected.  
However, we should also take account of another fact that the specific-heat experiment of CsCrF$_4$ captures no anomaly down to $1.5$K, while the bulk phase transition to the Cuboc phase should theoretically accompany a certain anomaly of the specific heat.
A reason for this inconsistency is an anisotropic interaction effect.
Since the inter-tube coupling $J_3$ of CsCrF$_4$ is basically very small, the Dzyaloshinsky-Moriya interaction\cite{DM_dzyaloshinsky, DM_moriya}, which is actually suggested in CsCrF$_4$ due to its crystal structure\cite{CsCrF4_Manaka_2011}, can compete with the small $J_3$ coupling.
Then, such an anisotropy effect may affect the Cuboc order configuration, and the transition with a small scale anomaly could be easily modified into a weak crossover.
A direct comparison of the neutron scattering experiment of CsCrF$_4$ with the Cuboc order is highly desired.
In addition to the above, we should also analyze how the quantum fluctuation affects the stability of the Cuboc order, which is another significant problem to understand the CsCrF$_4$ experiment.

In this paper, we have basically investigated the strong leg-coupling region ($J_1\ge J_2 \gg |J_3|$), since our motivation is in the spin tube system.
Our result implies that the leg-coupling certainly stabilizes the Cuboc order to be a true long-range order with the simultaneous transition of the spin and chirality.
On the other hand, the Cuboc order was originally proposed for the planar Kagom\'e model with the next-nearest-neighbor coupling, where the spin and $Z_2$ chirality transitions are separated\cite{cuboc_Domenge,domenge2008chirality}.
Recently, a Cuboc-type spin fluctuation was actually suggested for Kapellasite, which may be described by a $S=1/2$ Kagom\'e antiferromagnet containing up to third-nearest-neighbor couplings\cite{faak2012kapellasite}.
Theoretical investigations stimulated by Kapellasite also have revealed interesting properties attributed to the Kagom\'e structure.\cite{Janson_1,Janson_2,suttner2014renormalization,chiral_bieri}
Thus, it is an essential problem to understand how the 3D Cuboc class can be connected to the spin-liquid-like behavior with the $Z_2$-chirality breaking in the limit of the planar Kagom\'e-triangular model.

\begin{acknowledgements}

The authors would like to thank T. Okubo and K. Harada for fruitful discussions and comments. 
They are also grateful for H. Manaka, T. Masuda, and  M. Hagihara for discussions about experiments of CsCrF$_4$.
This work was supported in part by Grants-in-Aid No. 26400387 and 23340109 from the Ministry of Education, Culture, Sports, Science and Technology of Japan. 
\end{acknowledgements}

\end{document}